\documentclass{ws-procs9x6}

\begin{document}

\title{ICECUBE, THE WORLD'S LARGEST DARK MATTER DETECTOR }

\author{Hagar Landsman for the IceCube collaboration}

\address{Physics department, University of Wisconsin,\\
Madison, WI 53703, USA\\
$^*$E-mail: hagar@icecube.wisc.edu\\
http://icecube.wisc.edu}

\begin{abstract}
  IceCube is a kilometer scale high-energy neutrino
  observatory, currently under construction at the South Pole.
  It is a photo-detector, using the deep Antarctic ice as
  detection medium for the Cherenkov photons induced by relativistic 
  charged particles. These charged particles may be atmospheric muons
  or reaction products from neutrino interactions in the vicinity of
  the instrumented volume.
  The experiment searches for neutrinos originating in astrophysical
  sources, and can also detect neutrinos
  from WIMP interaction in the Sun or Earth.
  In the last two austral summers,  9 in-ice strings and 16 surface 
IceTop stations (out of up to 80 planned) were successfully deployed, and the 
detector has been taking data  ever since.
  In this proceedings, IceCube design, present status, performance and 
dark
  matter detection sensitivities will be discussed.

\end{abstract}


\bodymatter
\section{Introduction}\label{aba:sec1}
High energy neutrino detectors consist of an array of light detectors (``optical modules'') spread evenly in deep 
and clear media, like the Antarctic ice. They are designed to detect 
Cherenkov photons radiated by charged 
particles. The 
particles that are most likely to penetrate through the matter above the 
fiducial volume of the detector are muons (from interactions of cosmic rays in the atmosphere), and neutrinos (atmospheric or from any other source). 
Based on measurements of the number of photons arriving at different light detectors and their arrival times, a track 
or a cascade can be reconstructed. 

The IceCube detector is a neutrino detector currently under construction 
at the South 
Pole \cite{design}.
The main scientific goal of IceCube is to map the neutrino sky, 
searching for high energy neutrinos that may be created in hadronic 
processes, for example in Gamma-Ray Bursts or Active Galactic Nuclei. IceCube will also look for high energy GZK 
neutrinos 
\cite{GZK} that are expected to be created by interaction of the cosmic ray protons with the background radiation.
It will study air showers, and high energy atmospheric neutrinos \cite{atmospheric}. 
In a special data acquisition mode, IceCube will detect 
supernovae neutrinos, where the detection probability depends on the 
distance and neutrino flux from the explosion \cite{supernova}.
To some extent, IceCube measurements can 
be used 
for neutrino mass hierarchy and CP phase measurements \cite{winter}. 

Any model predicting a certain neutrino flux, can be tested by IceCube. 
These sources include, but are not limited to, dark matter, 
super-symmetry, magnetic monopoles and quantum gravity.  

The neutralino is the lightest Weakly Interacting Massive Particle 
(WIMP) 
candidate of the Minimal Super-Symmetric extension of the Standard Model (MSSM). 
It is stable due to R-parity conservation. 
In direct searches, the WIMPs are detected by the scattered nuclei in 
the rare elastic scattering of WIMPs on a suitable target \cite{cdms}. 
In indirect searches the annihilation products of two neutralinos are 
detected. If the decay products include 
neutrinos, neutrino telescopes like IceCube and  AMANDA can be used for detection \cite{wimp}. 

Relic neutralinos from the Galactic halo can lose energy through scattering 
with matter and become trapped in the gravitational field of massive objects, such as the Sun and the Earth, where they
accumulate over time and annihilate with each other.
The decay products may vary, and most of them will interact and decay in the massive body. If neutrinos are created 
from those secondaries, they will escape and create a neutrino flux.

The number of detected neutrinos, from neutralino interaction, depends on the source characteristic (like capture 
rate, neutralino annihilation cross section and mass) 
and on the detector response.

\section{WIMPs searches in AMANDA}
AMANDA, the predecessor experiment to IceCube, consists of 677 optical modules deployed on 19 strings at depth between 1.5 km and 2 km.
A search for an excess of muon neutrinos from the center of the Earth or the Sun was preformed using the AMANDA detector \cite{amanda-Sun,amanda-Earth}.

Muons created by cosmic-ray interactions in the atmosphere, can 
penterate through the ~1-2 km of ice and trigger the detector. Their 
abundance will overpower smaller fluxes. They are filtered using the 
matter of Earth, blocking muons from the northen hemisphere, but 
allowing neutrinos to pass through. Therefore only events transversing 
through Earth, ``upgoing events'' are used. 
For neutrinos with energies above one PeV, the interaction 
cross section is high enough and Earth starts to become opaque, 
depending on the angle the neutrino comes in.
 For WIMP studies, we are intrested in lower energies since
it is predicted that the mass of the neutralino will be less than 340 
TeV, based on unitarity conditions. 

Since the position of the Sun in the South Pole, reaches no more than 
$~23^\circ$ below the horizon, it is not possible to exploit the full 
advantage of the  upgoing events and the separation between signal and atmospheric muons is harder. In the case of searches for WIMPs from the center of the Earth, the expected signal is veritically upgoing, and the atmospheric muon background is easily removed, with no signal loss.

In both cases, tracks are reconstructed using times of hits in optical modules at known positions in the ice. The reconstructed zenith angle is used to separate background events from signal in the Earth analysis, and the space angle between the Sun and the reconstructed track is used to separate signal from background for the Sun analysis. 

The neutralino induced events were simulated using the DARKSUSY \cite{darksusy} program, with neutralino masses in the range of 100 GeV to 5 TeV. Two decay channels were analyzed: A 'soft' channel where the neutralinos annihilate to a pair of b-quarks, and a 'hard' channel where they decay to W-bosons. 

The background consists of atmospheric neutrinos that arrive to the detector 
from all directions (upgoing and downgoing events) 
and from mis-reconstructed downgoing muon. 
The background is estimated using off source measurements and 
simulation.  
No excess of signal was observed in either the Sun or the Earth 
analysis. 

\begin{figure}[h]
\begin{minipage}{13 pc}
\includegraphics[width=13 pc]{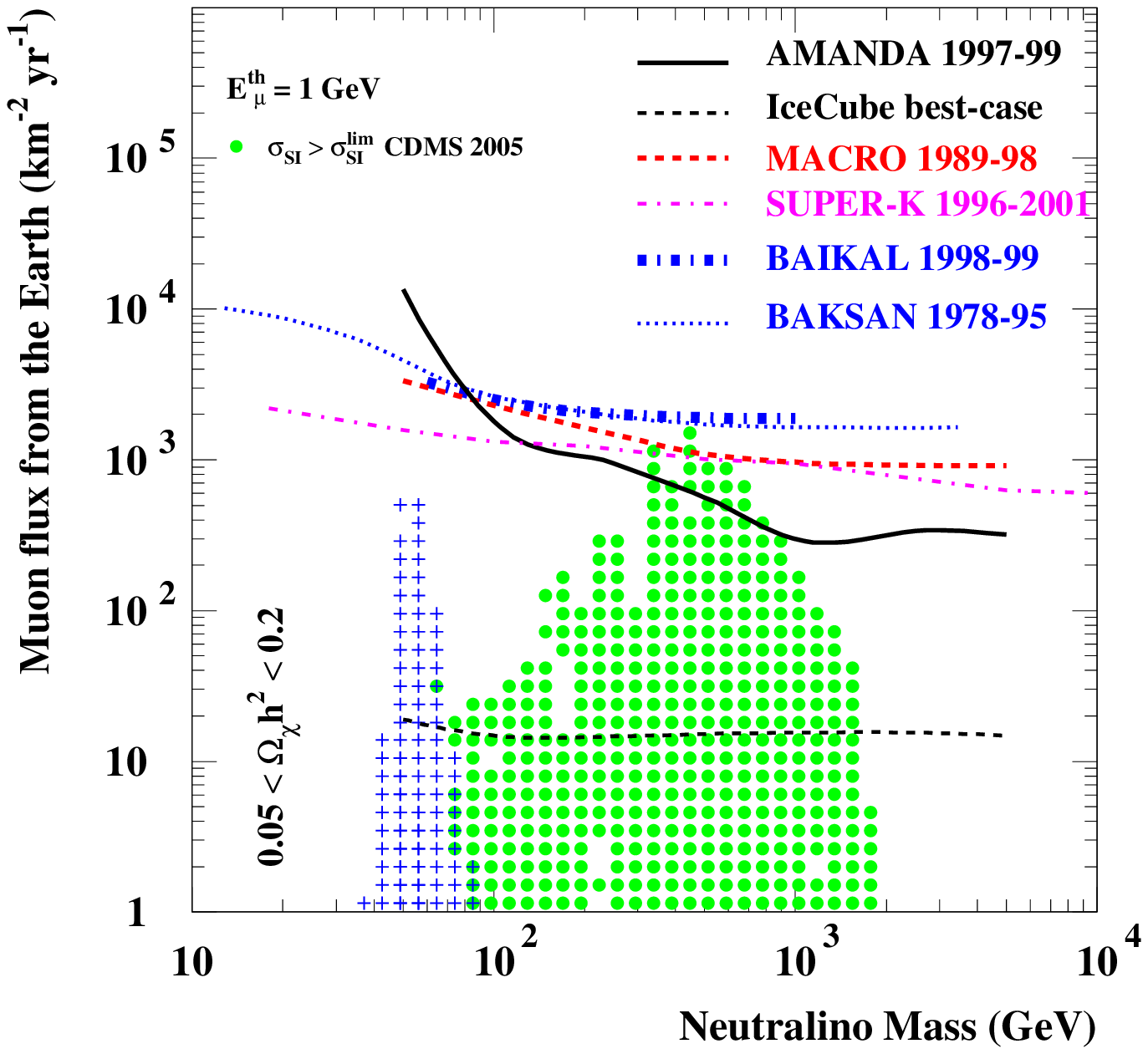}
\caption{\label{Earth-limits} 
Upper limits on the muon flux from the Earth, from neutralino 
annihilation 
in the 'hard' channel, as a function of the 
neutralino mass. 
IceCube best case refers to simulation of 80 strings with 3 
years worth of data. 
}
\end{minipage}\hspace{1pc}%
\begin{minipage}{13 pc}
\includegraphics[width=13 pc]{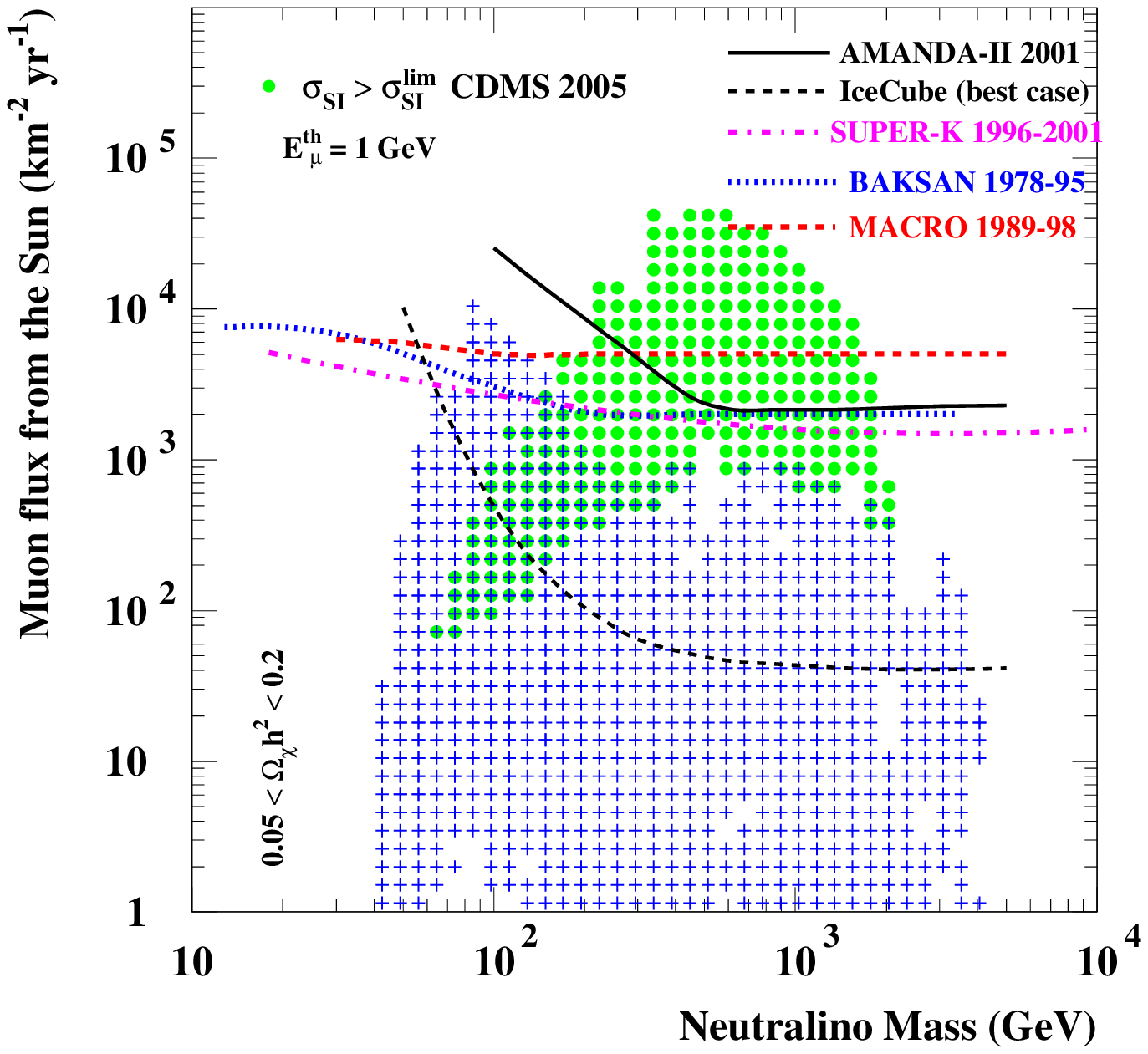}
\caption{\label{Sun-limits}
Upper limits on the muon flux from the Sun, from neutralino annihilation in the 'hard' channel, as a function of the 
neutralino mass. The crosses show predictions based on DARKSUSY. The 
dots show models disfavored by direct searches. 
The limits were rescaled to muon threshold of 1 GeV.}
\end{minipage} 
\end{figure}

Based on theoretical models and simulation this result is translated to 
upper 
limits on the flux of neutrinos originating from the Sun or from the 
Earth, and further extrapolated to limits on models. 
These predictions are model dependent. Not only they depend on SUSY  
models, that determine branching ratios, but to some extent also
on astrophysical assumptions.
The threshold energy of the detector plays an important role in the limit extrapolation. 
As can be seen in figures \ref{Earth-limits} and \ref{Sun-limits}, once 
completed, based on the 
current simulations, IceCube will contribute to the limits of WIMPs from the Sun.
 
\section{Current Status and Verification of IceCube}

A larger and improved detector will increase the number of events that 
can be measured and improve the track reconstruction resolution allowing 
better separation between signal and background. This will increase the 
sensitivity for WIMPs, and allow us to measure more neutralinos, in case 
of detection, or set better limits on models.

The IceCube detector will consist of up to 4800 Digital Optical Modules (DOMs) covering a fiducial volume of 1 cubic km \cite{design}.  The DOMs will be equally spaced on up to 80 strings at depths from ~1.5 to ~2.5 km in the clear Antarctic ice.  An array of surface stations, called IceTop, enhances the ability to trigger on, or veto, down-going showers. Each IceTop station consists of two clear ice tanks, each instrumented with 2 DOMs.  An IceTop station is located roughly 10 meters from each bore hole of the array.



Each DOM is an autonomous data collecting and digitizing unit consisting of a 10" Hamamatsu PMT in a 13" 
pressure sphere.
A main board inside the DOM can digitize up to 300 
Mega Samples per Second (MSPS) for 400 
ns and 40 MSPS for $6.4 \mu s$.  A flasher board, equipped with 
12 LEDs, produces 
pulses used for optical and timing calibration.


In the winter of 2004-2005 a single IceCube In-Ice string and 4 IceTop 
stations were deployed. At the end of the 2006 Austral summer IceCube 
consisted of 9 strings and 16 IceTop stations. A set of measurements was 
performed after the first year to confirm the design goal of the detector and check its performance \cite{firstyear}.

\subsection{Timing}
\begin{figure}[h]
\begin{minipage}{13 pc}
\includegraphics[width=13 pc]{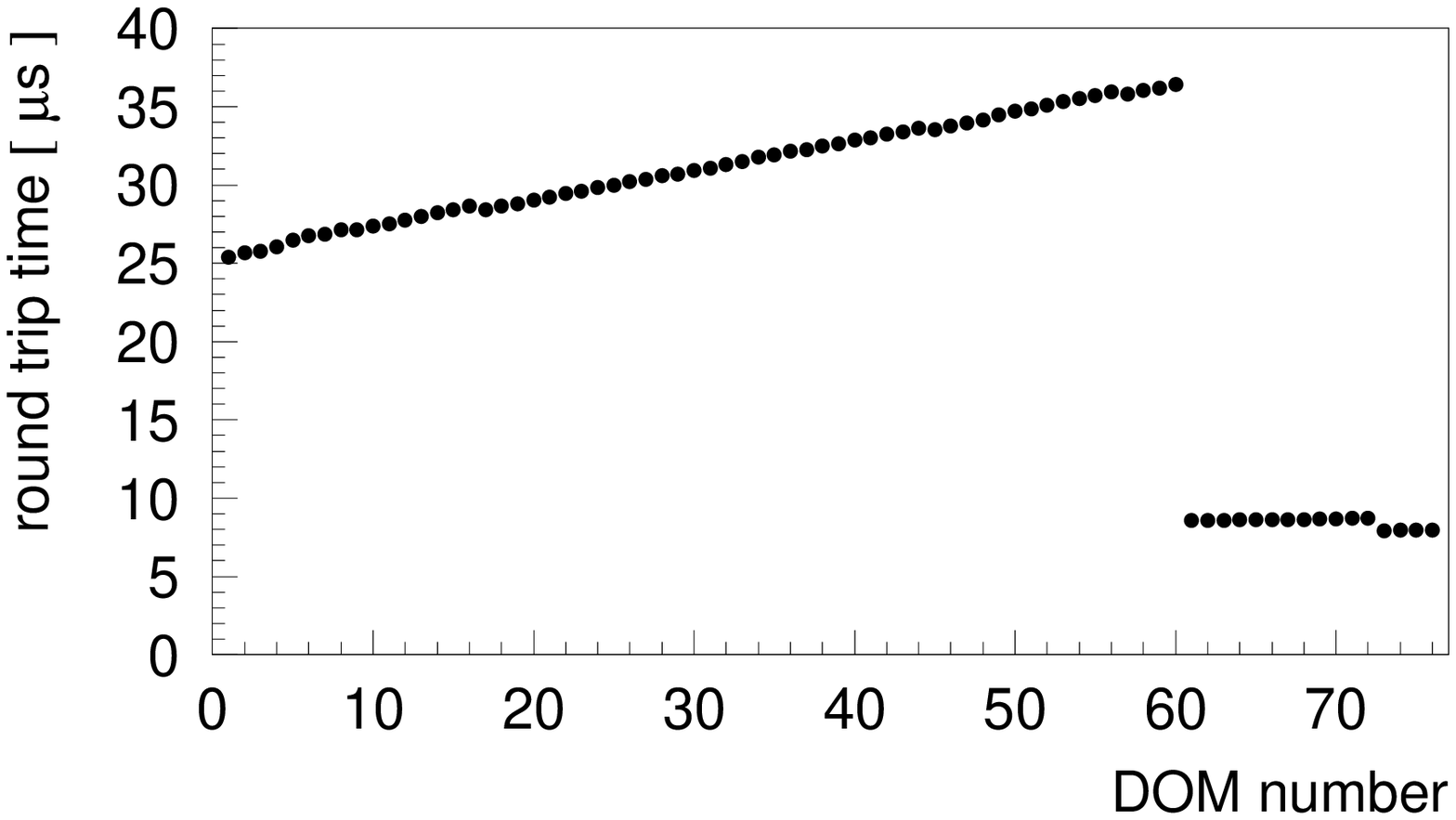}
\caption{\label{RoundTrip} Round trip time of the time calibration 
pulse. DOM number 1 is at the top of the string, and DOM 60 is at the 
bottom. DOMs 60-76 are part of the surface array. 
}
\end{minipage}\hspace{1pc}%
\begin{minipage}{13 pc}
\includegraphics[width=13 pc]{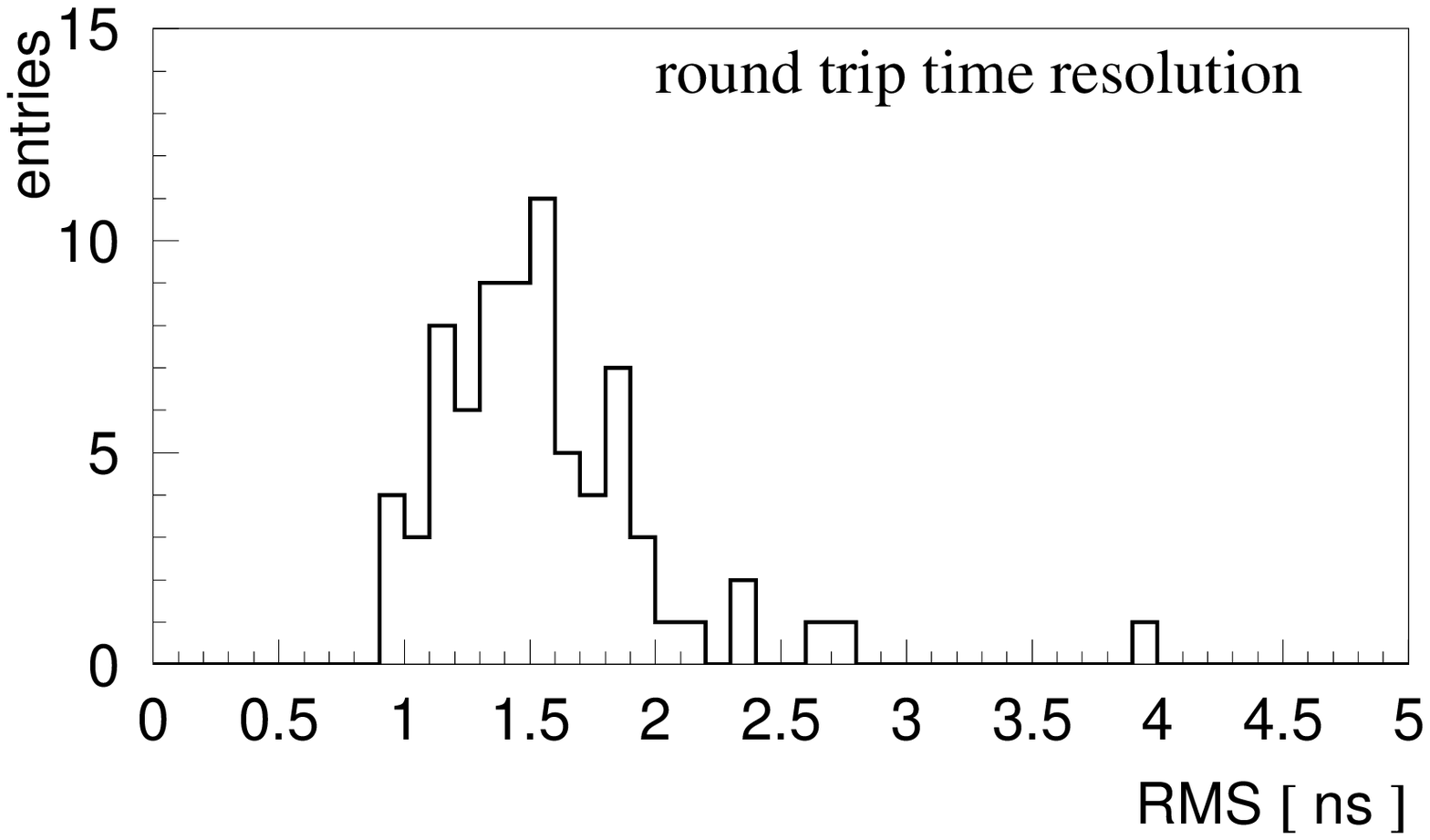}
\caption{\label{RapCALRMS}  The rms variation of the round trip time of the time calibration pulse for 76 DOMs on IceCube's first string and 4 IceTop tanks.
}
\end{minipage} 
\end{figure}

In order to reconstruct tracks over the entire array, a timing 
resolution of a few nanoseconds is needed. The time calibration of the entire array is done by sending a fast bipolar pulse at known intervals from the surface to every DOM. Each DOM then sends an identical fast bipolar pulse back to the surface after a known delay. Based on the time measured on the surface and on the DOM, and based on the shape of the measured pulse, the round trip time of the signal is calculated. Figure \ref{RoundTrip} shows the round trip times of the calibration pulse. The time increases with the depth of a DOM, and is constant for the IceTop array. 
The variation of the round trip time  from one calibration to the next 
provides a measurement of the time calibration accuracy (figure \ref{RapCALRMS}).

\begin{figure}[h]
\begin{minipage}{13 pc}
\includegraphics[width=13 pc]{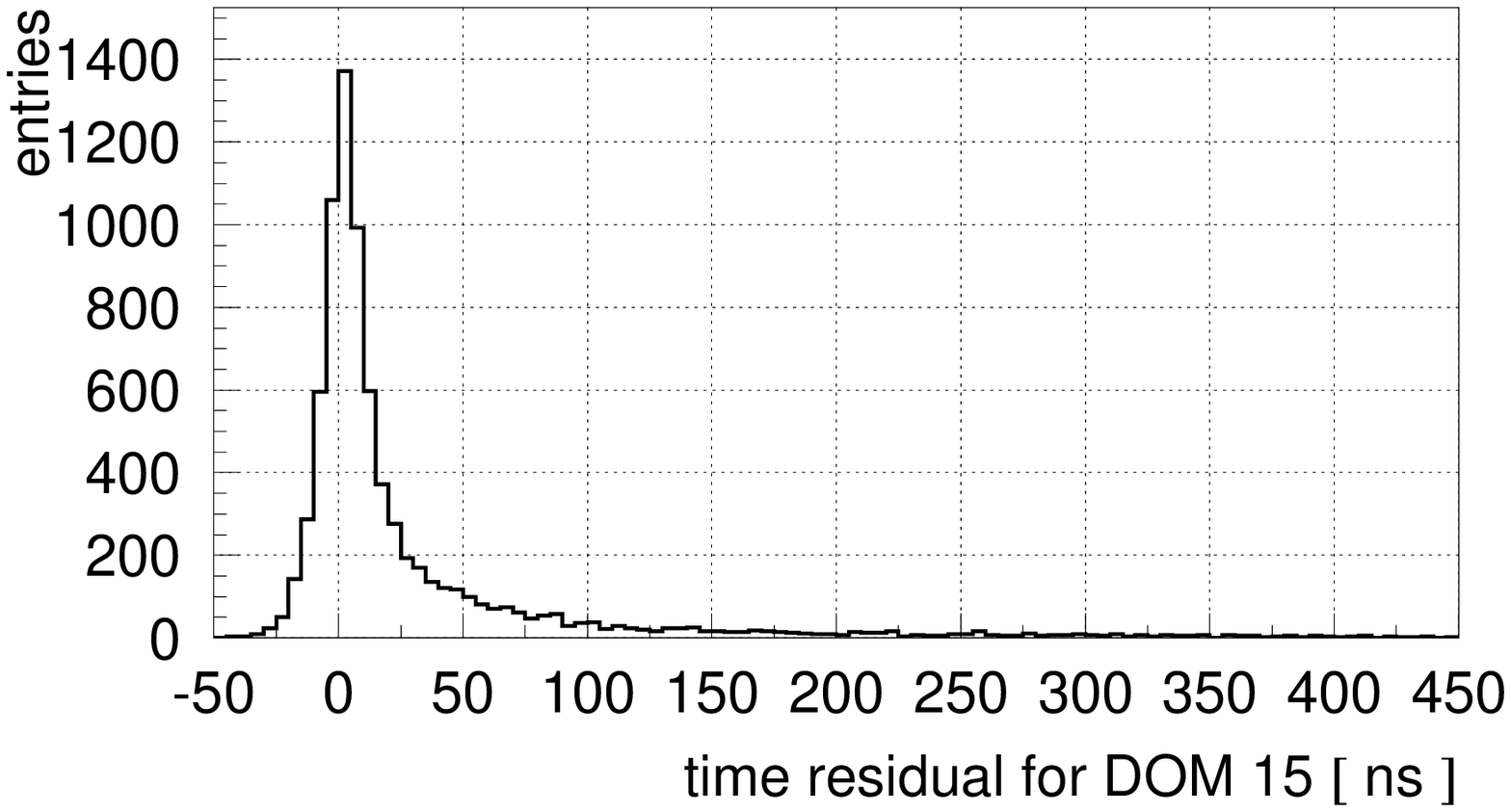}
\caption{\label{timeRes1}  Distribution of time residuals of
photons arriving at a DOM from nearby tracks reconstructed with the rest of the string.
}
\end{minipage}\hspace{1pc}%
\begin{minipage}{13 pc}
\includegraphics[width=13 pc]{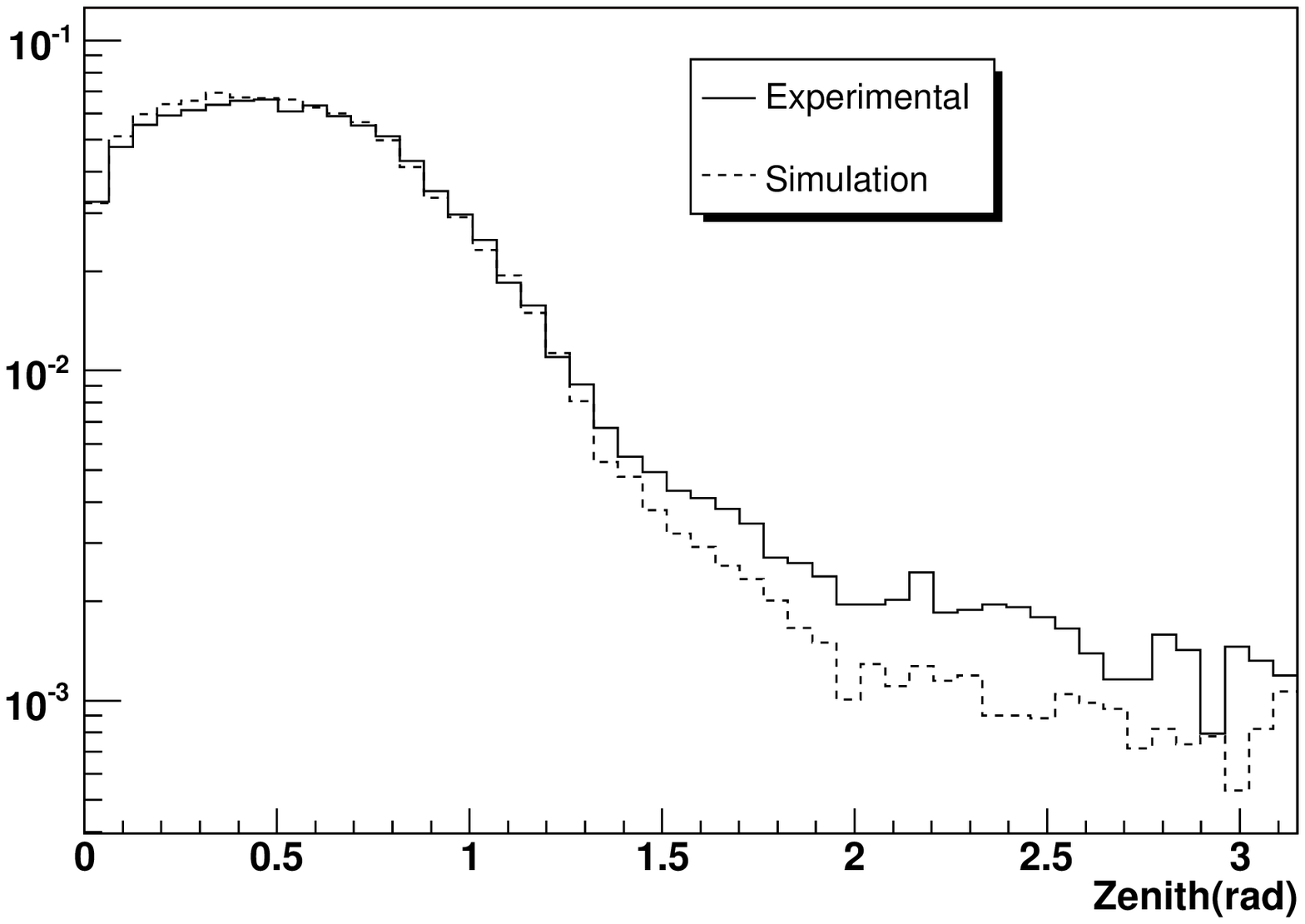}
\caption{\label{zenith}
Distribution of reconstructed and measured zenith angle for downgoing events.
}
\end{minipage} 
\end{figure}

The time resolution of each DOM was verified in two independent ways. First, a LED on a DOM was flashed and an adjacent DOMs triggered on it.  The time delay between the flashing and the triggering was measured multiple times. This procedure was repeated for all DOMs, and the maximum time RMS resolution was found to be less than 2ns.  
A different way to estimate the time resolution is by reconstructing down-going muon tracks excluding one DOM, and calculating the time difference between the measured hit time and the expected hit time. Figure \ref{timeRes1} shows the distribution of those time residuals for a single DOM using multiple events. The process is repeated for all DOMs. The resolution was found to be less than 3 ns, after accounting for ice properties.  


\subsection{Muon track reconstruction}

In order to point back to the origin of the neutrinos, a muon track is reconstructed and the zenith and azimuth angles are calculated. 
Reconstruction of muon tracks, originating from downward going cosmic rays, 
or from neutrino interactions in the detector volume, 
is done by analyzing the waveform measured on each DOM.

\begin{figure}[h]
\includegraphics[width=18 pc]{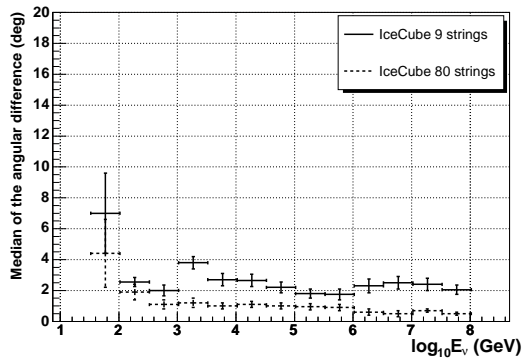}
\caption{\label{resolution-IC}
The angular resolution for 9 strings and for the full IceCube array as function of event energy (preliminary).  
}
\end{figure}

   A comparison of the measured atmospheric neutrinos and downgoing 
muons flux to the predicted and known fluxes is an important consistency 
check on the detector response and simulation.
The quality of the reconstruction depends on the number of DOMs hit 
(determined by the geometry and the energy of the event), the ice 
quality (impurities lead to photon scattering and absorption) and 
timing resolution. 
Figure \ref{zenith} compares the measured downgoing events reconstructed zenith angle to the predicted one using simulation of downgoing muons. 
Figure \ref{resolution-IC} shows the predicted angular resolution of the 
IceCube detector with 9 and 80 strings.

\subsection {Summary}
A neutrino flux from the center of the Earth or the Sun, may 
indicate neutralino capture inside those heavy bodies.
No excess of events from the Sun or the Earth, above the estimated 
atmospheric neutrino background, was measured by the AMANDA detector. Limits on the fluxes have been set 
for different neutralino masses.

The IceCube detector, under construction in the South Pole will be a 
powerful tool for high energy neutrino searches. Among other results, it 
will also improve the limits on neutrinos from neutralino annihilation, 
in the case of no signal. 
The higher threshold energy will decrease its sensitivity to 
lower mass neutralinos, but its larger size and better 
track reconstruction abilities, together with data collected on AMANDA, will improve the limits, especially for higher mass.
Until December 2007, IceCube consisted of 9 strings and 16 surface 
stations. The detector will rapidly grow as more strings and stations will 
be deployed. Measurements done on IceCube show good timing and 
reconstruction ability throughout the detector.


\begin{thebibliography}{9}
\bibitem{design} IceCube Project Preliminary Design Document, Ahrens et al. (IceCube Collaboration), http://icecube.wisc.edu.
\bibitem{GZK} R.Engel, D.Seckel and T.Stanev, Phys.Rev. D64 093010 (2001)
\bibitem{atmospheric} M.C. Gonzalez-Garcia, F.Halzen and M.Maltoni, Phys.Rev. D71, 092010 (2005)
\bibitem{supernova} J.Ahrens et al. (AMANDA Collaboration), Astropart. Phys. 16  , 345 (2002)
\bibitem{winter} W. Winter, Phys.Rev. D74, 033015 (2006)
\bibitem{cdms} CDMS Collaboration, Phys.Rev.Lett. 93 211301 (2004)
\bibitem{wimp} F. Halzen and D.Hooper, Phys. Rev. D73 1233507 (2006) 
\bibitem{amanda-Sun} A. Achterberg et al. (IceCube collaboration), Astropart. phys. 24 (2006) 459-466 
\bibitem{amanda-Earth} A. Achterberg et al. (IceCube collaboration), Astropart. Phys. 26 (2006) 129-139
\bibitem{darksusy} P.Gondolo et al., Journ. of Cosm. \& Astropart. Phys. 0407, 008, 2004
\bibitem{firstyear} A. Achterberg et al. (IceCube collaboration),  
Astropar. Phys. 26,  (2006) 155 

\end{thebibliography}
\end{document}